\documentclass[a4paper,
               biblatex,     
               keeplastbox,   
               ]{jacow}

\usepackage{pdfpages,multirow,ragged2e} %
\usepackage{tikz}
\usepackage{hyperref}
\usepackage{nameref}

\makeatletter%
	\ifboolexpr{bool{xetex}}
	 {\renewcommand{\Gin@extensions}{.pdf,%
	                    .png,.jpg,.bmp,.pict,.tif,.psd,.mac,.sga,.tga,.gif,%
	                    .eps,.ps,%
	                    }}{}
\makeatother

\ifboolexpr{bool{xetex} or bool{luatex}} 
 {}                                      
 {\usepackage[utf8]{inputenc}}           

\usepackage[USenglish]{babel}

\ifboolexpr{bool{jacowbiblatex}}%
 {%
  \addbibresource{Bibliography.bib}
 }{}
\newlist{inlinelist}{enumerate*}{1}
\setlist*[inlinelist,1]{%
	label=\textit{(\roman*)},
}
\usepackage{ccicons}
\usepackage{eso-pic}
\usepackage{graphicx}
\definecolor{colorMarge}{RGB}{255,255,255}
\newlength{\distance}
\newlength{\rulethickness}
\newlength{\ruleheight}
\newlength{\xoffset}
\newlength{\yoffset}
\AddToShipoutPicture{%
  \setlength\fboxsep{0pt}
  \ifodd\value{page}%
    \setlength{\xoffset}{\dimexpr\paperwidth-\rulethickness-\distance}%
  \else
    \setlength{\xoffset}{\distance}%
  \fi
  \AtPageLowerLeft{%
    \put(\LenToUnit{\xoffset},\LenToUnit{\yoffset}){%
      \colorbox{colorMarge}{%
        \parbox[b][\ruleheight][b]{\rulethickness}{%
          \centering

            \rotatebox[origin=lB]{90}{\small{\strut \quad\quad\quad \ccby \quad Content from this work may be used under the terms of the CC BY 3.0 licence (© 2022). Any distribution of this work must maintain attribution to the author(s), title of the work, publisher, and DOI}}%

        }%
      }%
    }%
  }%
}

\usepackage[absolute]{textpos}
\begin{document}

\begin{textblock}{10}(2,2)
\noindent\small 18th Int. Conf. on Acc. and Large Exp. Physics Control Systems\\
ISBN: 978-3-95450-221-9 \quad\quad\quad\quad\quad\quad\quad\quad ISSN: 2226-0358
\end{textblock}
\begin{textblock}{10}(11.4,2)
\noindent\small ICALEPCS2021, Shanghai, China \quad\quad JACoW Publishing\\
doi:10.18429/JACoW-ICALEPCS2021-MOPV042
\end{textblock}

\title{\NoCaseChange{PLCverif}: status of a formal verification tool for Programmable Logic Controller}

\author{I. D. Lopez-Miguel\thanks{ignacio.david.lopez.miguel@cern.ch}, J-C. Tournier, B. Fernandez, CERN, Geneva, Switzerland}
	
\maketitle

\begin{abstract}
Programmable Logic Controllers (PLC) are widely used for industrial automation including safety systems at CERN. The incorrect behaviour of the PLC control system logic can cause significant financial losses by damage of property or the environment or even injuries in some cases. Therefore ensuring their correct behaviour is essential. While testing has been for many years the traditional way of validating the PLC control system logic, CERN developed a model checking platform to go one step further and formally verify PLC logic.
This platform, called PLCverif, was first released internally for CERN usage in 2019, is now available to anyone since September 2020 via an open source licence. In this paper, we will first give an overview of the PLCverif platform capabilities before focusing on the improvements done since 2019 such as the larger support coverage of the Siemens PLC programming languages, the better support of the C Bounded Model Checker backend (CBMC) and the process of releasing PLCverif as an open-source software.
\end{abstract}

\section{Introduction}
Programmable Logic Controllers (PLC) are widely used for industrial automation including safety systems at CERN. The incorrect behaviour of the PLC control system logic can cause significant financial losses by damage of property or the environment or even injuries in some cases. Therefore ensuring their correct behaviour is essential. While testing has been for many years the traditional way of validating the PLC control system logic, it is often not sufficient as the sole verification method: testing, even when automated, can not be exhaustive, thus can not guarantee the correctness of a logic. Some types of requirements, such as safety (i.e. an unsafe state can never be reached) or invariant (formulas which shall be true over all possible system runs), can be very difficult, if not impossible, to test. Model checking is a formal verification technique which complements the testing activities in order to fully validate and verify a PLC control system logic. Model checking assesses the satisfaction of a formalised requirement on a mathematical model of the system under analysis. It checks the requirement's satisfaction with every input combination, with every possible execution trace. In addition, if a violation is found, a trace leading to the violated requirement is provided. The main hurdle to the widespread usage of model checking within the PLC community is twofold: (1) the mathematical model representing the system under analysis can be difficult to write and requires in-depth understanding of the model checking tools; and (2) many real-life PLC logics are too complex and face the state-space explosion problem, i.e. the number of possible input combinations and execution traces is too big to be exhaustively explored.

In 2019 CERN developed the PLCverif platform with the goals of easing the usage of model checking tools for the PLC developers community by automating the translation of the PLC programs to their mathematical models and to implement several abstraction algorithms to limit the state-space explosion problem. Since September 2020, the platform has been released under an open source license to foster the usage and the development of the tool within the PLC community. The objective of this paper is to give a status of the PLCverif platform focusing on the latest developments improving the usability and performance of the tool. 

The rest of the paper is organized as follows: Section \textit{\nameref{sec:overview}} gives an overview of PLCverif to better understand the scope and the architecture of the platform. Section~\textit{\nameref{sec:oss}} focuses on the open source release of PLCverif by describing the process of releasing the source code and presenting the code organization. Finally sections \textit{\nameref{sec:latest}} and \textit{\nameref{sec:ongoing}} present respectively the latest and ongoing developments.

\section{\NoCaseChange{PLCverif} Overview} \label{sec:overview}

This section gives an overview of the PLCverif platform \cite{blancovinuela:icalepcs2019} before presenting the latest developments.

 \subsection{Verification Workflow}
 
Out of the box, PLCverif offers a model checking workflow for the analysis of PLC programs. The verification workflow is shown in Figure \ref{fig:workflow-overview} and it has the following main steps:

\usetikzlibrary{shapes.geometric}
\usetikzlibrary{arrows.meta}
\usetikzlibrary{arrows}
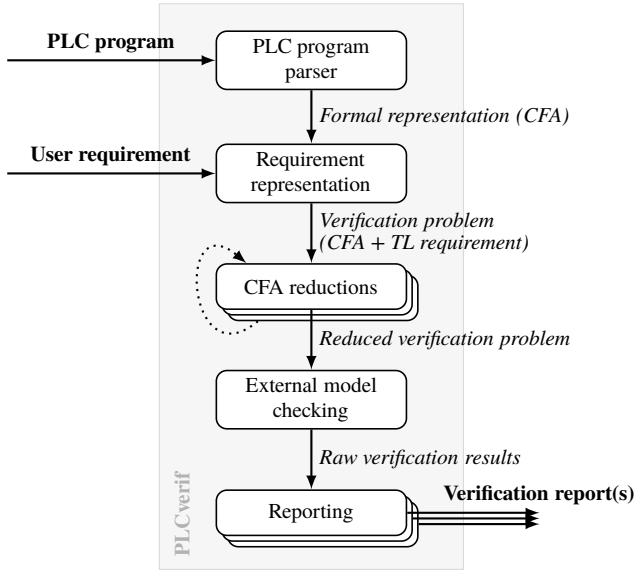
\begin{figure}[tb]
	\centering
	\footnotesize
	\begin{tikzpicture}
	
	\tikzstyle{box}=[rectangle,align=center,draw, semithick,minimum width=3cm,minimum height=0.6cm,rounded corners,fill=white]
	
	\tikzstyle{smallbox}=[box,minimum width=2.5cm]
	\tikzstyle{minibox}=[box,minimum width=1.5cm]
	\tikzstyle{arrow}=[-latex, thick]
	
	\filldraw[draw=black!15!white, fill=black!3.5!white] (-2,0.75) rectangle (2,-6.75);
	\node[align=left, rotate=90, text=black!30!white] at (-1.7,-6) {\textbf{PLCverif}};
	
	\node[smallbox] (parser) at (0,0) {PLC program\\parser};
	\node[smallbox] (reqrepr) at (0,-1.5) {Requirement\\representation};
	
	\node[smallbox] (reductions3) at (0.15,-3.15) {XXX};
	\node[smallbox] (reductions2) at (0.075,-3.075) {XXX};
	\node[smallbox] (reductions) at (0,-3) {CFA reductions};
	\node[smallbox] (mchecker) at (0,-4.5) {External model\\checking};	
	
	\node[smallbox] (reporting3) at (0.15,-6.15) {XXX};	
	\node[smallbox] (reporting2) at (0.075,-6.075) {XXX};	
	\node[smallbox] (reporting) at (0,-6) {Reporting};

	\draw[arrow](-4, 0)--(parser) node[midway,above,align=left]{\textbf{PLC program}};	
	\draw[arrow](-4, -1.5)--(reqrepr) node[midway,above,align=left]{\textbf{User requirement}};	
	
	\draw[arrow](parser)--(reqrepr) node[midway,right]{\textit{Formal representation (CFA)}};	
	\draw[arrow](reqrepr)--(reductions)
		node[midway,right,yshift=1.5mm]{\textit{Verification problem}}
		node[midway,right,yshift=-1.5mm]{\textit{(CFA + TL requirement)}};		
	\draw[arrow](reductions)--(mchecker) node[midway,right]{\textit{Reduced verification problem}};
	\draw[arrow](mchecker)--(reporting) node[midway,right]{\textit{Raw verification results}};
	\draw[arrow](reporting)--(3,-6) node[right,above]{\textbf{Verification report(s)}};
	\draw[arrow](reporting2)--(3,-6.075);		
	\draw[arrow](reporting3)--(3,-6.15);	
	\path (reductions3.200) edge [dotted,arrow,out=220,in=140,looseness=4] (reductions.160);
	\end{tikzpicture}
	\caption{Formal verification workflow of PLCverif}
	\label{fig:workflow-overview}
\end{figure}

\begin{enumerate}	
	\item \textbf{PLC program parsing.} PLCverif parses the PLC program (located in one or several files) to be analysed. By choosing the entry point of the verification, the analysis can be limited to a part of the program. The parsed PLC program is automatically translated into a mathematical, control flow-based representation, producing so-called Control Flow Automata (CFA). This precise description will serve as the basis for the analysis.
	
	\item \textbf{Requirement representation.} The user should describe the precise requirement to be checked. This, however, does not mean that the user needs to describe the requirement using mathematical formulae. Currently, two requirement description methods are supported:
		\begin{itemize}
			\item Assertion-based requirements: special comments in the source code (e.g. \texttt{//\#ASSERT On<>Off}) can describe expressions (invariants) which are expected to be always satisfied at a given location of the program. The verification job will then check if the violation of any of the selected assertions is possible.
			
			\item Pattern-based requirements: the user chooses a requirement pattern that is a precisely-phrased plain text sentence with some placeholders, e.g. ``\emph{If $\alpha$ is true at the end of the PLC cycle, then $\beta$ should always be true at the end of the same cycle.}''. The gaps in the requirement pattern ($\alpha$ and $\beta$ in the previous example) shall be filled with expressions over the PLC variables. For each requirement pattern, a defined temporal logic representation is defined which will be used in the next steps.
			
		\end{itemize}
	
	If needed, new types of requirement representations can be defined, adapted to the specific needs.
	
	\item \textbf{CFA reductions.} 
		The formal, precise CFA representation of the program, including also the requirement, may need to be reduced in order to make the verification feasible and efficient. These reductions will not change the verification result for the given requirement; however, they may remove parts of the program which do not influence the result of the currently checked requirement \cite{Darvas:FORTE2014}.

	\item \textbf{External model checking.}
		The model checking itself is performed by widely used model checker tools. In this step, 
		\begin{inlinelist}
			\item the reduced CFA is translated into the input syntax of the chosen model checker tool,
			\item the model checker tool is executed, and
			\item its output, notably the counterexample if available, is parsed to PLCverif's internal representation.
		\end{inlinelist}
	
		Currently the following external model checkers are supported: NuSMV \cite{Cimatti:CAV02}, nuXmv \cite{Cavada:CAV2014}, Theta \cite{Toth:FMCAD2017} and CBMC \cite{Clarke:TACAS2004}.
		These model checkers have different strengths and weaknesses. In addition, not every feature is supported by every model checker. 
		
	\item \textbf{Reporting.}
		The last step of the formal verification workflow is to produce verification reports. Some of these reports are in human-readable form and target the user of PLCverif. Other reports are machine-readable and serve as descriptions for the execution environment or as artifacts for later summary reports. 
\end{enumerate}

\section{Open Source Release} 
\label{sec:oss}

PLCverif is available publicly since September 2020 along with its source code under an EPL-2.0\footnote{\label{noteEPL}\url{https://www.eclipse.org/legal/epl-2.0/}} license on GitLab\footnote{\url{https://gitlab.com/plcverif-oss}}. This section presents the reasons for open sourcing PLCverif, as well as the process to choose the open source license and the code organization as found in the GitLab repository.

\subsection{Motivations}
PLCverif has been entirely developed at CERN and therefore fits CERN's needs. Nevertheless, the platform could be beneficial to two different communities outside CERN: the PLC developers community and the model checker community. For PLC developers, PLCverif could be used out of the box if the language used is among the ones already supported, i.e. Siemens Statement List (STL) and Siemens Structure Control Language (SCL). The current coverage\footnote{Coverage was calculated as the number of instructions that are implemented in PLCverif vs. the number of instructions in a given grammar for a given PLC. Binary arithmetic operations like addition or subtraction, and binary logic operations like conjuntion or disjuntion are not taken into account when the coverage for SCL is calculated. That is the main reason why the figures for SCL are much lower than for STL.} of Siemens STL and SCL in PLCverif is currently at 66\% and 40\%, respectively, for S7-300/400 PLCs, and at 55\% and 25\%, respectively, for S7-1200/1500 PLCs. This covers most of the functionalities of PLC programs developed at CERN as the instructions implemented represent the core of those languages. However, support for a missing instruction could be easily added if needed. Similarly, the support of a new programming language could also be developed taking as a reference the current implementation for the Siemens languages. For the model checking community, PLCverif offers the possibility to be integrated in a platform verifying real-life PLC code: it is a great opportunity for this community to test new algorithms or improvements of existing model checkers. 

\subsection{License Selection}
From the different open source licenses available, the choice has been made to release the PLCverif platform source code under the Eclipse Public License 2.0 (EPL\footnote{See footnote \ref{noteEPL}.}). This license is similar to the GNU General Public License (GPL\footnote{\url{https://www.gnu.org/licenses/gpl-3.0.en.html}}) but allows to link the code to proprietary applications: it then allows to use and extend the tool, even for commercial purposes. The reason for choosing the EPL-2.0 license was driven by the fact that is a weak copyleft license, but also that most of the components used by PLCverif are already under the EPL-2.0 (more details in \cite{blancovinuela:icalepcs2019}).

\subsection{GitLab Repository}
The PLCverif platform has been developed within the Eclipse ecosystem: it is mainly written in Java (Java 11), Xtend and Xtext. The modularity is assured via the split of the code among several distinct projects and Eclipse plugins. The main projects of the platform are: 
\begin{itemize}
    \item PLCverif Backend\footnote{\url{https://gitlab.com/plcverif-oss/cern.plcverif}} is the core of the PLCverif platform and is responsible for all the CFA manipulations represented in Figure \ref{fig:workflow-overview}. It is also responsible for interacting with the external model checkers via dedicated Eclipse plugins. 
    \item Siemens Support\footnote{\url{https://gitlab.com/plcverif-oss/cern.plcverif.plc.step7}} is responsible for the parsing of the Siemens STL and SCL code, i.e. transforms the PLC user code into the PLCverif internal CFA representation. 
    \item PLCverif Frontends\footnote{\url{https://gitlab.com/plcverif-oss/cern.plcverif.gui}}$^{,}$\footnote{\url{https://gitlab.com/plcverif-oss/cern.plcverif.cli}} are the visible parts from a user point of view. The GUI project provides a graphical application embedding a PLC code editor, a specification requirement editor and a report visualization part. The CLI project is the way to execute the PLCverif workflow via the command line allowing the execution of headless verification jobs such as in a CI/CD (Continuous Integration / Continuous Deployment) pipeline for PLC code.  
\end{itemize}

\section{Latest developments}
\label{sec:latest}
Since the publication of \cite{blancovinuela:icalepcs2019} in 2019, several improvements have been made. Some of them are summarized below:

\begin{itemize}
    \item The C code used to run CBMC was originally produced directly converting the intermediate model to a C code using \textit{goto} instructions. CBMC is however not efficient with this kind of programs since it is not able to find loops. This method has been changed in order to produce a structured C code without \textit{gotos}, being able to efficiently use the option \textit{--partial-loops} of CBMC. With this option, CBMC will execute loops only partially. The disadvantage of this option is that the counterexample might be spurious.
    
    \item In order to confirm the feasibility of a counterexample produced by PLCverif when a property is violated, it is common to try to reproduce that situation in a real PLC or via simulation. In order to automate this process, it is now possible to automatically generate a file with the values of all the variables that can be used as an input to the simulator or to the real PLC.
    
    \item Safety programs in Siemens are written in function block diagram (FBD) language. After exporting them with \textit{OpennessScripter}\footnote{TIA Portal Openness API.}, an XML file is produced. A new feature in PLCverif has been developed in order to import those XML files into PLCverif translating them into STL code.
    
    \item The coverage of Siemens programs was increased both for STL and SCL. More built-in functions from TIA portal were included.
    
    \item Support of latest Theta version was included. Theta is being actively developed and new releases have been launched. In order to keep up with the latest improvements, PLCverif has been adapted to correctly parse Theta output. Currently, PLCverif supports Theta v2.21.0.
    
    \item The intermediate model Control Flow Instance had the limitation that it could not be used with dynamic indexing arrays. However, Theta supports this feature and PLCverif has been adapted to generate Theta programs with dynamic indexes.
    
    \item The grammar implemented in PLCverif to parse Siemens PLC programs has been extended to include partial support of Schneider PLC programs.
    
    \item Upgrade to Java 11. PLCverif was originally developed in Java 8. However, in order not to lose support and to be able to run the latest versions of some model checkers (Theta), it was needed to upgrade to Java 11.
\end{itemize}

\section{On-going Challenges and Developments}
\label{sec:ongoing}

\subsection{Simplification of numeric variables}

As observed in the large code base of CERN industrial PLC systems, one of the main challenges to perform PLC model checking is the state-space explosion originated by the inclusion of numeric variables. PLCverif represents input variables as non-deterministic in the intermediate model. This means that a 16-bit integer is going to have $2^{16}\approxeq 7 \cdot 10^4$ possible values that the model checker needs to explore.

There exist some techniques to handle this type of variables, such as counterexample-guided-abstraction refinement (CEGAR) \cite{clarke:cegar} or Satisfiability Modulo Theories (SMT). However, other approaches to directly simplify the PLCverif intermediate model are under investigation, highly improving the performance of the NuSMV model checker \cite{lopez:memocode21}.

\subsection{Iterative verification} 

It is common practice to have different modules within PLC projects (see \cite{Fernandez:ICALEPCS2021} for an example). Some of these projects are too large to be verified by PLCverif yet. However, if the program is split into parts, the verification cases are smaller and can be successfully executed. If it would be possible to combine all the results together, it would be feasible to verify these large programs. To this end, different compositional verification approaches have been analysed but no generic method has been found yet to be applied to PLC projects.

Nonetheless, other abstraction techniques are under consideration, such as abstracting away the different modules. With this approach, a verification case is executed with all the functions abstracted away (all their outputs are going to be non-deterministic). If the property is satisfied, the original program satisfies that property too. On the other hand, if it is violated, one needs to check if the abstracted functions can produce the outputs leading to the violation. If it is not possible, a function is concretized (it comes back to its original form) and a new verification case is executed. This process is continued until a feasible counterexample is found or until all functions are concretized (coming back to the original program). Different strategies and improvements can be investigated for this method.

\subsection{Counterexample analysis}

When a program is verified and the result is a violation of the property, a counterexample is given by PLCverif. If the program is composed of several variables and they interact with each other (see \cite{Fernandez:ICALEPCS2021} for an example), the counterexample is going to be large. Therefore, it will be difficult to analyse what part of the code made the property fail.

Some investigations have been done in this direction in order to point the user to possible locations in the code that have an impact on the final assertion. This way, the user would not need to go through all the code but just focus on the highlighted parts.

\subsection{Other} 

Since the release of PLCverif, there has been some progress in the development of more efficient model checkers. As already explained previously, the latest version of Theta has been integrated into PLCverif. However, although CBMC is efficient, it is a SAT-based model checker that uses few abstraction techniques. Therefore, an SMT-based model checker like ESBMC \cite{Mikhail:ASE2018} could improve the performance of CBMC.

\section{Conclusion}
\label{sec:conclusion}
This paper presented the latest developments of the PLCverif platform to formally verify PLC programs. The developments can be grouped into two main lines of work:
\begin{itemize}
    \item Promoting and easing the use of PLCverif by making it open source, by supporting more PLC manufacturers (i.e. Schneider Electric), and by guiding the user to the root cause of an issue when a property is violated (counterexample analysis).
    \item Improving the performance of the verification time by simplifying numerical variables without loosing meaningful information and by implementing an iterative verification process allowing to verify even more complex applications.
\end{itemize}

All the different developments presented in this paper are in different stages of maturity and are in the pipeline to be included into PLCverif. In addition some new developments will be carried on to support Schneider safety programs and to integrate new model checkers such as ESBMC.


%
%
\ifboolexpr{bool{jacowbiblatex}}%
{\printbibliography}%

%
%


\end{document}